\documentclass[english,aps,pra,showpacs,twocolumn,groupedaddress,floatfix]{revtex4}

\usepackage{amsmath}
\usepackage{amssymb}
\usepackage{amsfonts}
\usepackage{bbm}
\usepackage{epsfig}
\usepackage{babel}
\usepackage{color}

\usepackage{mathtools}
\usepackage{textcomp}

\newcommand{\ud}{\mathrm{d}}
\newcommand{\Id}{\operatorname{id}}

\newcommand{\Tr}{\operatorname{Tr}}
\newcommand{\bra}{\langle}
\newcommand{\ket}{\rangle}
\newcommand{\mc}[1]{\mathcal{#1}}
\newcommand{\pdag}{{\phantom{\dag}}}

\newcommand  {\veps}{\varepsilon}
\renewcommand{\L}{\mc{L}}
\newcommand  {\K}{\mc{K}}

\renewcommand{\H}{\mc{H}}
\newcommand{\Z}{\mc{Z}}

\newcommand{\V}{\mc{V}}
\newcommand{\B}{\mc{B}}
\newcommand{\NN}{\mathbb{N}}
\newcommand{\RR}{\mathbb{R}}
\newcommand{\LL}{\mathbb{L}}
\newcommand{\dm}{\rho}

\newcommand{\diam}{\operatorname{diam}}
\newcommand{\norm}[1]{\left\Vert #1\right\Vert}
\newcommand{\normS}[1]{\Vert #1\Vert}
\newcommand{\normt}[1]{\boldsymbol{\vert} #1\boldsymbol{\vert}}

\newcommand{\hl   }[1]{{\bar{#1}}}
\newcommand{\Latt}{\Lambda}
\newcommand{\compl}[1]{{\Latt\setminus #1}}
\newcommand{\vol}  [1]{{\operatorname{Vol}(#1)}}
\newcommand{\HS} {{\operatorname{HS}}}
\newcommand{\InfToInf}{$(\infty\!\to\!\infty)$}
\newcommand{\OneToOne}{$(1\!\to\!1)$}

\newtheorem{theorem}{Theorem}

\definecolor{grey}{rgb}{.5,.5,.5}

\newcommand{\appx}{Appx.}

\newcommand{\fu} {Dahlem Center for Complex Quantum Systems, Freie Universit{\"a}t Berlin, 14195 Berlin, Germany}
\newcommand{\potsdam} {Institute for Physics and Astronomy, University of Potsdam, 14476 Potsdam, Germany}

\newcommand{\Title} {Quasi-locality and efficient simulation of Markovian quantum dynamics}
\newcommand{\Authors}
{
\author{Thomas Barthel}
\author{Martin Kliesch}
\affiliation{\fu}
\affiliation{\potsdam}
}
\newcommand{\Date} {February 12, 2012}

\begin{document}

\title{\Title}
\Authors

\begin{abstract}
We consider open many-body systems governed by a time-dependent quantum master equation with short-range interactions. With a generalized Lieb-Robinson bound, we show that the evolution in this very generic framework is \emph{quasi-local}, i.e., the evolution of observables can be approximated by implementing the dynamics only in a vicinity of the observables' support.
The precision increases exponentially with the diameter of the considered subsystem.
Hence, the time-evolution can be simulated on classical computers with a cost that is independent of the system size. Providing error bounds for Trotter decompositions, we conclude that the simulation on a quantum computer is additionally efficient in time. For experiments and simulations, our result can be used to rigorously bound finite-size effects.
\end{abstract}


\date{\Date}

\maketitle

\section{Introduction}
In Lorentz-invariant theories, a maximum speed for the propagation of information is, by construction, the speed of light. In nonrelativistic quantum theory, the existence of a maximum propagation speed results more indirectly and for different reasons. For nonpathological models, this maximum speed is much smaller than the speed of light. The seminal paper by Lieb and Robinson \cite{Lieb1972-28} and further contributions like \cite{Bratteli1997-2,Hastings2004-69,Hastings2006-265,Nachtergaele2006-265,Bravyi2006-97,Eisert2006-97,Osborne2006-97,Nachtergaele2007-12a,Burrell2007-99,Burrell2009-80,Nachtergaele2009-286,Schuch2011-84} cover isolated systems.

Here, we consider the evolution of a more general and, experimentally, extremely relevant class of systems -- open quantum many-body systems governed by a quantum master equation \cite{Davis1976,Alicki2007} with short-range Liouvillians that are allowed to be time-dependent. Prominent experimental examples are presented in Refs.~\cite{Myatt2000-403,Viola12001-293,Deleglise2008,Barreiro2010-6,Barreiro2011-470}, and recent theoretical advances on quantum computation, nonequilibrium steady states, and phase transitions in open systems can, for example, be found in Refs.~\cite{Diehl2008-4,Verstraete2009-5,Prosen2011-107a,Prosen2011-107b}.
Going beyond the existence of a finite maximum propagation speed and the existence of a well-defined thermodynamic limit \cite{Lieb1972-28,Nachtergaele2011-552}, we show that the time-evolution of such systems is \emph{quasi-local}. This means that, up to an exponentially small error, the diameter of the support of any evolved local observable grows at most linearly in time, or, put differently, that the evolution of the local observable can be approximated to arbitrary precision by applying the propagator of a spatially truncated version of the Liouvillian; Fig.~\ref{fig:LRtrotter}b. For the special case of isolated systems, where the evolution is given by a unitary transformation, the corresponding question has been addressed in Ref.~\cite{Nachtergaele2007-12a}. As a tool for the proof of quasi-locality, we derive and employ a Lieb-Robinson-type bound very similar to the recent results of Poulin \cite{Poulin2010-104} and Nachtergaele \emph{et al.} \cite{Nachtergaele2011-552}. All constants in the bounds are given explicitly in terms of the system parameters.

The quasi-locality of Markovian quantum dynamics has several crucial consequences. It implies that the evolution of observables with a finite spatial support can be simulated efficiently on classical computers, in the sense that the computation cost is independent of the system size, irrespective of the desired accuracy. This can for example be exploited in an exact diagonalization approach for a sufficiently large vicinity of the support of the considered observable; Fig.~\ref{fig:LRtrotter}b. For more sophisticated simulation techniques, we provide, in extension of Ref.~\cite{Kliesch2011-107}, error bounds for Trotter decompositions \cite{Trotter1959} of the subsystem propagator into a circuit of local channels; see Fig.~\ref{fig:LRtrotter}c. The Trotter error is polynomial in the time, at most linear in the size of the time step, and can hence be made arbitrarily small. Importantly, the subsystem Trotter decompositions allow for the efficient simulation of the time-evolution on a quantum computer as envisaged by Feynman. For any required accuracy, the simulation can be implemented with a cost that is independent of the system size and polynomial in the time.

Experimental and numerical physicists who study nonequilibrium systems can use our result on quasi-locality to rigorously bound finite-size effects. This is for example relevant for experiments with ultracold atoms in optical lattices \cite{Bloch2007} and numerical investigations employing time-dependent density-matrix renormalization group methods \cite{Vidal2003-10,White2004,Daley2004,Schollwoeck2005}.

\section{Setting}\label{sec:setting}
\subsection{Lattice and equations of motion}
Let us consider lattice systems, where each site $z\in\Latt$ is associated with a local Hilbert space $\H_z$. Subsystem Hilbert spaces are denoted by
\begin{equation*}
	\H_V \coloneqq \bigotimes_{z \in V}\H_z\,\, \forall_{V\subset \Latt}\quad\text{and}\quad
	\H \coloneqq \H_\Latt.
\end{equation*}
Let $\dm(t)$ denote the system state at time $t$. Markovian dynamics of an open quantum system, i.e., the evolution under a linear differential equation that generates a completely positive and trace-preserving map for $\dm$, can always be written in the form of a \emph{Lindblad equation} \cite{Lindblad1976-48,Gorini1976-17,Wolf2008-279}
\begin{equation*}
	\partial_t\dm=-i[H,\dm]+\sum_\nu\left( L_\nu^\pdag\dm L_\nu^\dag -\frac{1}{2}(L_\nu^\dag L_\nu^\pdag\dm+\dm L_\nu^\dag L_\nu^\pdag)\right),
\end{equation*}
where the arbitrary \emph{Lindblad operators} $L_\nu$ and the Hermitian Hamiltonian $H$ may depend on time.
This equation captures, for example in the framework of the \emph{Born-Markov approximation}, the evolution of a system that interacts with an environment \cite{Davis1976,Alicki2007} and isolated systems as a special case.
Let us switch from the Schr\"{o}dinger picture, where expectation values are evaluated according to $\bra O \ket_{s\to t}=\Tr[ \dm(t) O]$ with $\dm(s)=\dm$, to the Heisenberg picture, where $\bra O \ket_{s\to t}=\Tr[ \dm O(s)]$ with $O(t)=O$. The corresponding time-dependence of an observable $O(s)\in\B(\H)$ is then given by the quantum master equation
\begin{equation*}
	\partial_s O(s)=-\L(s)O(s),
\end{equation*}
where $\L(t)\in \B(\B(\H))$ is a \emph{super-operator}, the so-called \emph{Liouvillian}, with the Lindblad representation
\begin{equation*}
	\L O = i[H,O]+\sum_\nu\left( L_\nu^\dag O L_\nu^\pdag -\frac{1}{2}(L_\nu^\dag L_\nu^\pdag O+O  L_\nu^\dag L_\nu^\pdag)\right).
\end{equation*}
The set of Liouvillians with spatial support $V\subset\Latt$ will be denoted by $\LL_V\subset \B(\B(\H_V))$.

\subsection{Short-range Liouvillian}
In order to be able to use Lieb-Robinson bound techniques, we need to restrict ourselves to Liouvillians with norm-bounded short-range interaction terms.
Let us hence assume that $\L$ is a sum of local Liouville terms $\ell_Z$ with norm bound $\normt{\ell}$, maximum range $a$, and a maximum number $\Z$ of nearest neighbors \footnote{The results of this article follow similarly for systems with long-range interactions of sufficiently fast decay. For the sake of readability we refrain from presenting this more general scenario.}. Specifically,
\begin{align}\label{eq:Liouvillian}
	\L(t) &= \sum_{Z\subset \Lambda} \ell_Z(t),\quad \ell_Z(t)\in\LL_Z,\\\label{eq:normBound}
	\normt{\ell} &\coloneqq \sup_{t,Z\subset\Lambda}\norm{\ell_Z(t)},\\
	a &\coloneqq \sup_{Z:\ell_Z\neq 0}\diam(Z),\\\label{eq:NoNeighbors}
	\Z &\coloneqq \max_{Z:\ell_Z\neq 0}|\{Z'\subset\Lambda\,|\,\ell_{Z'}\neq 0,\, Z'\cap Z\neq\varnothing\}|,
\end{align}
where $\diam(Z) \coloneqq \max_{x,y\in Z}d(x,y)$ is the diameter of $Z$ and $d$ is a metric on the lattice $\Lambda$.
In Eq.~\eqref{eq:normBound}, we have used the super-operator norm defined by $\norm{T}:=\sup_{O\in\B(\H)} \norm{T O}/\norm{ O }$. In the Heisenberg picture, this is the physically relevant norm as induced by the operator norm $\norm{ O }$; see Appx.~\ref{sec:norms}.
For notational convenience, we define for every subsystem $V\subset\Latt$ the corresponding extension $\hl{V}$, volume $\vol{V}$, and truncated Liouvillian $\L_V$,
\begin{align}\label{eq:LiouvillCl}
	\hl{V}  &\coloneqq\textstyle \bigcup_{\underset{Z\cap V\neq \varnothing}{Z:\ell_Z\neq 0}} Z,\\
	\vol{V} &\coloneqq |\{Z\subset V\,|\, \ell_Z\neq 0 \}|,\\
	\L_V(t) &\coloneqq\textstyle \sum_{Z\subset V} \ell_Z(t).
\end{align}

\subsection{Propagators}
\emph{Propagators} $\tau_V(s,t)$ are super-operators that map observables to time-evolved observables. They are defined as the unique solutions of
\begin{equation}\label{eq:propDerive1}
	\partial_s \tau_V(s,t) = -\L_V(s) \tau_V(s,t),\quad \tau_V(t,t)=\Id \quad\forall_{s\leq t}.
\end{equation}
With $\tau(s,t) \coloneqq \tau_\Latt(s,t)$ one has indeed $O(s)= \tau(s,t) O(t)$. Propagators obey the composition rule $\tau(r,s)\tau(s,t)=\tau(r,t)$ $\forall_{r\leq s\leq t}$.
As discussed in Appx.~\ref{sec:propagators},
the derivative with respect to the second time argument is given by
\begin{equation}\label{eq:propDerive2}
	\partial_t \tau_V(s,t) =  \tau_V(s,t)\L_V(t),
\end{equation}
and the propagators are norm-decreasing,
\begin{equation}\label{eq:propNorm}
	\norm{\tau(s,t)O}\leq \norm{O}\quad \forall\,{\L\in\LL_\Latt,\,s\leq t,\, O\in\B(\H)}.
\end{equation}

\section{Quasi-locality of the evolution}
\begin{figure*}[t]
\centering
\includegraphics[width=0.975\linewidth]{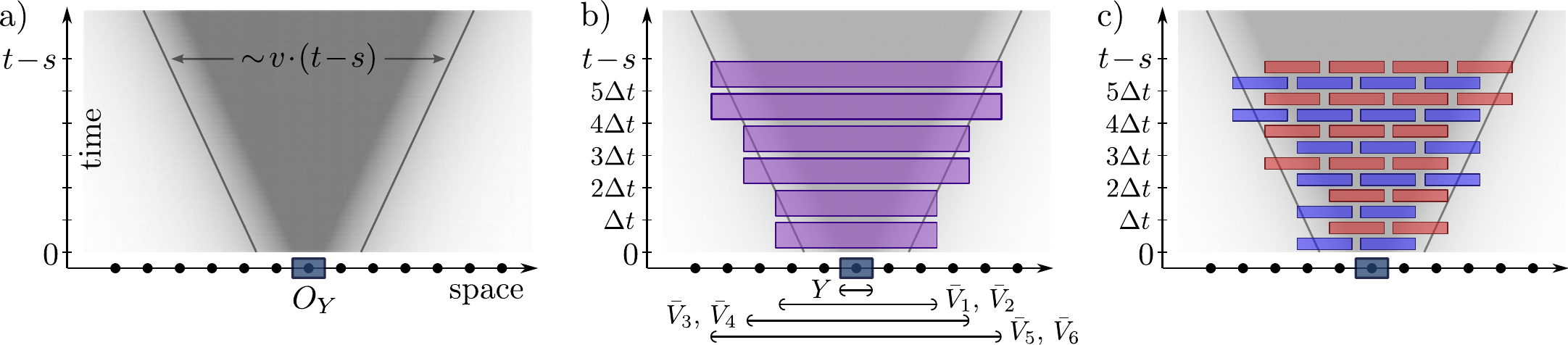}
\caption{
(a) An evolved local operator $\tau(s,t)O_Y$ behaves almost like the identity outside its associated space-time cone.
(b) Approximating $\tau(s,t)O_Y$ by application of subsystem propagators to $O_Y$. The errors decrease exponentially with the subsystem sizes.
(c) For one-dimensional systems, approximating $\tau(s,t)O_Y$ by a Trotter decomposition yields an error scaling as $(t-s)^2\Delta t$. Note that the Trotter circuit can be trimmed off at the boundary of the Lieb-Robinson space-time cone.}
\label{fig:LRtrotter}
\end{figure*}
Given an operator $O_Y\in\B(\H_Y)$ with support $Y\subset \Latt$, we would like to show that the exactly time-evolved operator $\tau(r,t)O_Y$ with $r\leq t$ can be approximated by the evolution  with respect to a spatially truncated Liouvillian, i.e., by $\tau_\hl{V}(r,t)O_Y$ with $Y\subset V\subset \Latt$. Indeed, our main result, Theorem~\ref{th:localEvol}, states that the approximation error is exponentially small, in the distance of $\compl{V}$ to the time-$r$ slice of a space-time cone originating from the operator's support $Y$ at time $t$; see Fig.~\ref{fig:LRtrotter}b. More precisely, the error decays exponentially in $d(Y,\compl{V})/a-v\cdot(t-r)$, where $d(X,Y) \coloneqq \inf_{x\in X,y\in Y}d(x,y)$ is the distance of two subsystems $X,Y\subset\Latt$, and $v=e\Z\normt{\ell}$ is the so-called Lieb-Robinson velocity.

To prove this, we can write the difference of the evolved operators in the form
\begin{align*}
	&\tau(r,t)O_Y - \tau_\hl{V}(r,t)O_Y\\
	&\quad= -\int_r^t\ud s\, \partial_s\left[ \tau_\hl{V}(r,s) \tau(s,t) \right]O_Y\\
	&\quad= \int_r^t\ud s\, \tau_\hl{V}(r,s)\underbrace{[\L(s)-\L_\hl{V}(s)]}_{=\L_\compl{V}(s)} \tau(s,t) O_Y
\end{align*}
due to the fundamental theorem of calculus and Eqs.~\eqref{eq:propDerive1} and \eqref{eq:propDerive2}.
Using the triangle inequality and the fact that the propagators are norm-decreasing, it follows that
\begin{align}\nonumber
	&\norm{\tau(r,t)O_Y - \tau_\hl{V}(r,t)O_Y}\\\label{eq:localApprox1}
	&\quad\leq \sum_{X\subset\compl{V}}
	   \int_r^t\ud s\, \norm{\ell_X(s)\tau(s,t) O_Y}.
\end{align}
In the case of unitary dynamics ($\ell_X(s)O=i[h_X,O]$), the integrand would be of the form 
$\norm{[h_X,\tau(s,t) O_Y]}$, and the standard Lieb-Robinson bound \cite{Lieb1972-28,Bratteli1997-2,Hastings2004-69,Hastings2006-265,Nachtergaele2006-265} would be applicable. To proceed in our more general case, however, we use a Lieb-Robinson bound for Markovian quantum dynamics, similar to recent results in Refs.~\cite{Poulin2010-104,Nachtergaele2011-552}.

\begin{theorem}\textbf{(Lieb-Robinson bound for Markovian quantum dynamics)}\label{th:LR}\\
Let the Liouvillian $\L(t)=\sum_{Z\subset \Lambda} \ell_Z(t)$ for the lattice $\Lambda$ be of finite range $a$, with a finite maximum number $\Z$ of nearest neighbors, and $\normt{\ell}$ as defined in Sect.~\ref{sec:setting}. Also, let $\K_X\in\LL_X$, $O_Y\in\B(\H_Y)$, and $r\leq t\in\RR$. Then
\begin{equation}\label{eq:LR}
	\norm{\K_X\tau(r,t)O_Y}\leq \V_{X,Y}\norm{\K_X}\norm{O_Y}\,e^{v\cdot(t-r)-d(X,Y)/a},
\end{equation}
where $v \coloneqq e\Z\normt{\ell}$ and $\V_{X,Y} \coloneqq \min\{\frac{\vol{\hl{X}}}{\Z},\frac{\vol{\hl{Y}}}{\Z}\}$.
\end{theorem}
The proof is given in Sect.~\ref{sec:LRbound}.
The theorem tells us that an evolved observable $\tau(r,t)O_Y$ remains basically unchanged when we evolve it with respect to a Liouvillian that is supported at a distance $R\gg v\cdot (t-r)$ away from $Y$, i.e., that $\tau(r,t)O_Y$ behaves like the identity outside the corresponding space-time cone.
In the special case $\K_X O=i[O_X,O]$, Eq.~\eqref{eq:LR} yields a Lieb-Robinson bound for $\norm{[O_X,\tau(r,t)O_Y]}$ as in Ref.~\cite{Poulin2010-104}.

This theorem can now be employed to proceed from Eq.~\eqref{eq:localApprox1} in our proof of quasi-locality. Let us restrict ourselves to the typical case of Liouvillians $\L(t)$ for which the number of terms $\ell_X(t)$ with distance ${d(y,X)}/{a}\in[n,n+1)$ from any site $y\in\Latt$ is bounded by a power law,
\begin{align}\label{eq:NoTerms}
	&|R_{n,y}|\leq M n^\kappa\quad\forall_{y\in\Latt,\,n\in\NN_+},\\\nonumber
	&\textstyle R_{n,y}\coloneqq\{X\subset\Lambda\,|\,\ell_{X}\neq 0,\, \frac{d(y,X)}{a}\in[n,n+1)\},
\end{align}
for some constants $M,\kappa>0$.
Now, choose a point $y_0\in Y$ that is closest to $\compl{V}$, i.e., $d(y_0,\compl{V})=d(Y,\compl{V})$.
With $D \coloneqq \lceil d(Y,\compl{V})/a\rceil$, we can exploit that the support of every term in $\L_\compl{V}$ is element of exactly one of the sets $R_{n,y_0}$ with $n\geq D$, to obtain
\begin{align*}
	&\norm{\tau(r,t)O_Y - \tau_\hl{V}(r,t)O_Y}\\
	&\quad\leq \sum_{n=D}^\infty\sum_{X\in R_{n,y_0}}
	   \int_r^t\ud s\, \norm{\ell_X(s)\tau(s,t) O_Y}\\
	&\quad\leq \sum_{n=D}^\infty M n^\kappa \normt{\ell}\norm{O_Y}
	   \int_r^t\ud s\, e^{v\cdot(t-r)-n}\\
	&\quad\leq M \normt{\ell}\norm{O_Y} \frac{e^{v\cdot(t-r)}}{v} \sum_{n=D}^\infty n^\kappa e^{-n}.
\end{align*}
In the second step, Theorem~\ref{th:LR} and $\V_{XY}\leq\vol{\hl{X}}/\Z\leq 1$ have been used.
With the bound $\sum_{n=D}^\infty n^\kappa e^{-n} \leq 2e D^\kappa e^{-D}$ $\forall_{D>2\kappa+1}$ from Appx.~\ref{sec:expBound}, we arrive at the central result of this work:

\begin{theorem}\textbf{(Quasi-locality of Markovian quantum dynamics)}\label{th:localEvol}\\
Let the Liouvillian $\L(t)=\sum_{Z\subset \Lambda} \ell_Z(t)$ for the lattice $\Lambda$ be of finite range $a$, with a finite maximum number $\Z$ of nearest neighbors, and $\normt{\ell}$ as defined in Sect.~\ref{sec:setting}. Further, let constraint Eq.~\eqref{eq:NoTerms} be fulfilled for some constants $M,\kappa>0$. Also, let $Y\subset V\subset\Latt$, $O_Y\in\B(\H_Y)$, and $r\leq t\in\RR$. Then one has with $D \coloneqq \lceil d(Y,\compl{V})/a\rceil$
\begin{multline}\label{eq:localEvol}
	\norm{\tau(r,t)O_Y - \tau_\hl{V}(r,t)O_Y}\\\textstyle
	\quad \leq \frac{2M}{\Z}\norm{O_Y} D^\kappa e^{v\cdot(t-r)-D}
	\quad\forall_{D>2\kappa+1},
\end{multline}
where $v$ is the Lieb-Robinson speed from Eq.~\eqref{eq:LR}.
\end{theorem}
The full dynamics can be approximated with exponential accuracy by subsystem dynamics.
In a sense, the constraint Eq.~\eqref{eq:NoTerms} requires the lattice to have a finite spatial dimension.
A $\mc{D}$-dimensional hypercubic lattice with finite-range interactions fulfills Eq.~\eqref{eq:NoTerms} with $\kappa = \mc{D}-1$.
An interesting observation is that short-range models on a \emph{Bethe lattice}
\cite{Bethe1935-150} have a finite Lieb-Robinson speed according to Theorem~\ref{th:LR} but do not fulfill Eq.~\eqref{eq:NoTerms} and are thus not covered by Theorem~\ref{th:localEvol}. For such systems, it is, hence, conceivable that a quench of the Liouvillian starting at time $t=0$ with a distance of at least $aD$ from some point $y$ causes a perceptible effect at $y$ for a time $t^*\ll D/v$.

\section{Trotter decomposition of the evolution}\label{sec:Trotter}
The quasi-locality of the dynamics, Theorem~\ref{th:localEvol}, implies that the evolution of observables with a finite spatial support can be simulated efficiently on classical computers, in the sense that the computation cost is independent of the system size, irrespective of the desired accuracy. However, exploiting this in an exact diagonalization approach that stores the approximated time-evolved observable $\tau_\hl{V}(r,t)O_Y$ in a full basis of $\H_\hl{V}$ exactly, requires resources that are exponential in the size $|\hl{V}|$ of the considered subsystem. There are more sophisticated numerical techniques, e.g., one can use matrix-product operators \cite{Zwolak2004-93,McCulloch2007-10,Hartmann2009-102} for the representation of (an approximation to) $\tau_\hl{V}(r,t)O_Y$ or sampling algorithms.
In such schemes, it is typically not possible to address the differential equation for $\tau_\hl{V}(r,t)O_Y$ directly, but one can use Trotter decompositions \cite{Trotter1959} instead, where propagators $\tau_\hl{V}(r,t)$ are decomposed into a circuit of local (diameter-$a$) channels. 

Using the quasi-locality, Theorem~\ref{th:localEvol}, and techniques as in Ref.~\cite{Kliesch2011-107}, we can derive a Trotter decomposition with an error that is polynomial in time, at most linear in the time step, and, in extension of Ref.~\cite{Kliesch2011-107}, system-size independent. Furthermore, implementing such a Trotter circuit on a quantum computer \cite{Kliesch2011-107} yields a simulation that, additionally to being independent of the system size, is efficient in time.
In this case, the physically relevant norm for super-operators $T$ is the subsystem-seminorm
\begin{equation}\label{eq:normSubsys}
	\norm{T}_Y \coloneqq \sup_{O_Y\in\B(\H_Y)}\norm{T O_Y}/\norm{O_Y}.
\end{equation}

\begin{theorem}\textbf{(Efficient Trotter decomposition of time-evolved observables)}\label{th:TrotterEvol}\\
With the preconditions of Theorem~\ref{th:localEvol}, a sequence of times $t_0\leq t_1\leq \dots \leq t_N$ and a sequence of subsystems $Y\subset V_1\subset V_2\subset \dots \subset V_N\subset\Latt$ such that $D_n \coloneqq  \lceil d(Y,\compl{V_n})/a\rceil >2\kappa+1$ $\forall_n$, the Trotter decomposition
\begin{equation}\label{eq:TrotterDecomp}
	\tilde{\tau} \coloneqq \prod_{n=1}^N \prod_{Z\subset \hl{V}_n:\,\ell_Z\neq 0} \tau_Z(t_{n-1},t_n)
\end{equation}
into propagators $\tau_Z$ for local Liouville terms $\ell_Z$ approximates the 
full system propagator $\tau(t_0,t_N)$ up to an error
\begin{align}\nonumber
	&\norm{\tau(t_0,t_N) - \tilde{\tau} }_Y
	\leq \sum_{n=1}^N ( {\textstyle\frac{2M}{\Z}} D_n^\kappa e^{v\cdot(t_n-t_0)-D_n} + \veps_n ),\\\label{eq:TrotterEvol}
	&\veps_n \coloneqq (t_n-t_{n-1})^2\Z \vol{\hl{V}_n} \normt{\ell}^2 e^{(t_n-t_{n-1})\normt{\ell}}
\end{align}
with the Lieb-Robinson speed $v$ from Eq.~\eqref{eq:LR}.
\end{theorem}
In the Trotter decomposition $\tilde{\tau}$, we used the convention $\prod_{n=1}^N T_n=T_1T_2 \dots T_N$, and the ordering of the channels $\tau_Z$ in the second product of Eq.~\eqref{eq:TrotterDecomp} can be chosen arbitrarily. As in Ref.~\cite{Kliesch2011-107}, one can use averaged Liouvillians, i.e., $\tau_Z(r,t)\mapsto e^{\int_r^t\ud s \ell_Z(s)}$, without changing the scaling of the error bound. Choosing a constant time step, $t_n=n\Delta t$, and subsystems $V_n$ such that $D_n= D_0 + v n\Delta t$, for sufficiently large $D_0$, the bound \eqref{eq:TrotterEvol} is dominated by the Trotter errors $\veps_n$. The subsystems can be chosen such that $\diam{V_n}\leq \diam(Y)+a D_n$; see Fig.~\ref{fig:LRtrotter}c. For this case, the total error is in $\mc{O}\left(\Delta t (\diam(Y)/a+D_0+ v t)^{\kappa+2} \right)$. Higher-order Trotter-Suzuki decompositions \cite{Suzuki1985-26} can be used to further improve the scaling in $\Delta t$.

To prove Theorem~\ref{th:TrotterEvol}, one can first apply Theorem~\ref{th:localEvol}, the inequality
$\normS{T_1T_2-\tilde{T}_1\tilde{T}_2}
\leq \norm{T_1}\normS{T_2-\tilde{T}_2} + \normS{T_1-\tilde{T}_1}\normS{\tilde{T}_2}$, and Eq.~\eqref{eq:propNorm} iteratively $N$ times, to obtain
\begin{equation}\label{eq:LR-Trotter}
	\normS{\tau(t_0,t_N) - \tau^V }_Y
	\leq {\textstyle\frac{2M}{\Z}} \sum_{n=1}^N D_n^\kappa e^{v\cdot(t_n-t_0)-D_n}
\end{equation}
with $\tau^V \coloneqq \prod_{n=1}^N\tau_{\hl{V}_n}(t_{n-1},t_n)$. For every time-step propagator $\tau_{\hl{V}_n}(t_{n-1},t_n)$, we can then employ a Trotter decomposition similar to Ref.~\cite{Kliesch2011-107}, yielding
\begin{multline}\label{eq:slice-Trotter}
	\normS{\tau_{\hl{V}}(r,t)
	-\prod_{Z\subset \hl{V},\,\ell_Z\neq 0} \tau_Z(r,t)}_Y\\
	\leq (t-r)^2\Z \vol{\hl{V}} \normt{\ell}^2 e^{(t-r)\normt{\ell}}.
\end{multline}
See Appx.~\ref{sec:doTrotter} for details. Combining Eqs.~\eqref{eq:LR-Trotter} and \eqref{eq:slice-Trotter} with the triangle inequality proves Theorem~\ref{th:TrotterEvol}.

\section{Proof of Theorem~\ref{th:LR}}\label{sec:LRbound}
With an argument similar to those in Refs.~\cite{Lieb1972-28,Bratteli1997-2,Hastings2004-69,Hastings2006-265,Nachtergaele2006-265,Poulin2010-104,Nachtergaele2011-552}, we want to bound the norm of the operator
\begin{equation}
	G(r) \coloneqq \K_X\tau(r,t)O_Y
\end{equation}
under the preconditions of Theorem~\ref{th:LR}.
$G$ is the solution to the final value problem $G(t)=\K_X O_Y$,
\begin{align*}\nonumber
	\partial_r G(r)&=-\K_X \L(r)\tau(r,t) O_Y\\
	&= -\L_\compl{X}(r)G(r) - \K_X \L_\hl{X}(r)\tau(r,t) O_Y,
\end{align*}
due to Eq.~\eqref{eq:propDerive1}, $\L=\L_\hl{X} + \L_\compl{X}$[Eq.~\eqref{eq:LiouvillCl}], and $\K_X \L_\compl{X}=\L_\compl{X}\K_X$ for all Liouvillians $\K_X\in\LL_X$.
As can be checked by differentiation, a corresponding integral equation for $G(r)$ is
\begin{multline*}
	G(r)=\tau_\compl{X}(r,t) G(t)\\
	  + \int_r^t\ud s\, \tau_\compl{X}(r,s)\K_X\L_\hl{X}(s)\tau(s,t)O_Y.
\end{multline*}
Using the triangle inequality, the norm-submulti\-pli\-ca\-ti\-vi\-ty, and the fact that the propagators are norm-decreasing, this yields the bound
\begin{align}\nonumber
	\norm{G(r)}
	 &\leq \norm{G(t)}+\norm{\K_X}\int_r^t\ud s\, \norm{\L_\hl{X}(s)\tau(s,t)O_Y}\\\label{eq:Gbound}
	 &\leq \norm{G(t)}+\norm{\K_X}\sum_{Z\subset\hl{X}} \int_r^t\ud s\, \norm{\ell_Z(s)\tau(s,t)O_Y}.
\end{align}
Now a Picard iteration for the related quantity
\begin{equation}\label{eq:defC}
	C_X(r) \coloneqq \sup_{\K\in\LL_X}\frac{\norm{\K\tau(r,t)O_Y}}{\norm{\K}}
\end{equation}
can be used to obtain a bound for $\norm{G(r)}$. Inserting Eq.~\eqref{eq:Gbound} in Eq.~\eqref{eq:defC} gives
\begin{align}\nonumber
	C_X(r)&\leq C_X(t)+\sum_{Z\subset\hl{X}} \sup_{s\in[r,t]}\norm{\ell_Z(s)} \int_r^t\ud s\, C_Z(s),\\\label{eq:C0}
	C_X(t)&\leq \delta(X,Y)\norm{O_Y},
\end{align}
where $\delta(X,Y)=1$ for $X\cap Y\neq\varnothing$ and $\delta(X,Y)=0$, otherwise. The second line follows from $\K_X O_Y=0$ for Liouvillians $\K_X\in\LL_X$ with $X\cap Y=\varnothing$, and $\norm{\K O_Y}\leq\norm{\K}\norm{O_Y}$ in general.
\begin{figure}[!tb]
\centering
\includegraphics[width=0.55\linewidth]{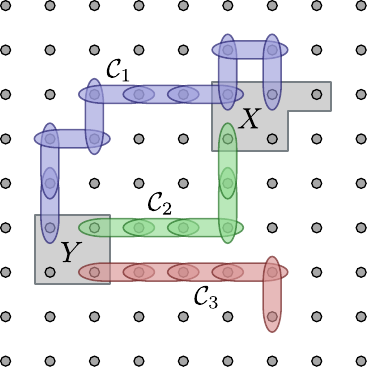}
\caption{In the proof of Theorem~\ref{th:LR}, we need to bound a sum over all paths of length $n$ starting in $X$, the support of $\K_X$, and ending in $Y$, the support of $O_Y$; see Eq.~\eqref{eq:C-sum}. A path corresponds to a sequence of local Liouville terms $(\ell_{Z_i})$ with overlapping supports. In the depicted situation of a two-dimensional lattice with nearest-neighbor interaction, path $\mc{C}_1$ would contribute to the sum for $n=10$ and $\mc{C}_2$ for $n=5$. To simplify the calculation for upper bounds, the sums are extended to contain \emph{all} paths starting in $Y$ (if $\vol{\hl{Y}}<\vol{\hl{X}}$; all paths starting from $X$, otherwise). Hence, in the bound for $n=5$, also paths like $\mc{C}_3$ are taken into account.}
\label{fig:LRproof}
\end{figure}
Starting the Picard iteration for $C_X(r)$ with Eq.~\eqref{eq:C0} and $Z_0 \coloneqq X$ leads to
\begin{gather}\label{eq:C-sum}
	C_X(r)\leq \norm{O_Y}\sum_{n=0}^\infty \frac{(t-r)^n}{n!}\,c_n\text{ with}\\\nonumber
	c_n=\sum_{Z_1\subset\hl{Z}_0}\sum_{Z_2\subset\hl{Z}_1}\hspace{-0.1em}\dots
	 \hspace{-0.3em}\sum_{Z_n\subset\hl{Z}_{n-1}}\hspace{-0.7em}\delta(Z_n,Y)
	 \prod_{i=1}^n\sup_{s\in[r,t]}\norm{\ell_{Z_i}(s)}.
\end{gather}
Now we can exploit that the Liouville terms are of finite range $a$ and that they induce a finite maximum number $\Z=\max_{Z:\ell_Z\neq 0}\vol{\hl{Z}}$ of nearest neighbors; Sect.~\ref{sec:setting}.
The sum in Eq.~\eqref{eq:C-sum} runs over all paths from $X$ to $Y$. Depending on whether $\vol{\hl{X}}$ or $\vol{\hl{Y}}$ is larger, the number of such paths with length $n$ can be bounded by the number of all length-$n$ paths starting in $X$ or $Y$, respectively. See Fig.~\ref{fig:LRproof}.
This gives the simple bound
\begin{align}\nonumber
	c_n&\leq \V_{X,Y}(\mc Z\normt{\ell})^n,\text{ and thus},\\\label{eq:C-exp-sum}
	C_X(t)&\leq\V_{X,Y}\norm{O_Y}\sum_{n=D}^\infty \frac{\theta^n}{n!},
\end{align}
where $\theta \coloneqq (t-r)\Z\normt{\ell}$, $D=\lceil d(X,Y)/a\rceil$, and $\Z\V_{X,Y}$ is the minimum of the numbers of Liouville terms $\ell_{Z_i}$ supported in $X$ and $Y$, $\V_{X,Y}=\min\{\vol{\hl{X}},\vol{\hl{Y}}\}/\Z$. We have also used that $c_n=0$ for all $n<D$, as one needs at least $D$ Liouville terms of overlapping support to pass from the subsystem $X$ to subsystem $Y$, such that $\delta(Z_n,Y)\neq 0$.
Using induction, the sum in Eq.~\eqref{eq:C-exp-sum} can be bounded by $\sum_{n=D}^\infty \frac{\theta^n}{n!} \leq e^{\theta e-D}$; see \appx~\ref{sec:expPartialBound}. Hence, Theorem~\ref{th:LR} follows,
\begin{equation*}
	\norm{G(t)}\leq \V_{X,Y}\norm{\K_X}\norm{O_Y}\, e^{(t-r)\Z\normt{\ell} e-D}.
\end{equation*}

\section{Conclusion}
We have shown that the evolution of an observable with support $Y$ under a quantum master equation with a short-range Liouvillian can be approximated by the evolution with respect to the truncation of the Liouvillian to a subsystem $V\supset Y$. The error decreases exponentially in the distance of $Y$ from the complement of $V$. With this tool, we derived an error bound for Trotter decompositions of the propagator. Those results correspond to efficient simulation techniques for open-system dynamics on classical and quantum computers and provide rigorous bounds to finite-size effects.

\acknowledgments
We gratefully acknowledge inspiring discussions with J.~Eisert, C.~Gogolin, V.~Nesme, and T.~J.\ Osborne. This work has been supported by the EU (Minos, Qessence), the EURYI, the BMBF (QuOReP), and the
Studienstiftung des Deutschen Volkes.

\appendix

\section{Operator and super-operator norms}\label{sec:norms}
In this work, two of the \emph{Schatten $p$-norms} \cite{Bhatia1997} are employed.
The \emph{$\infty$-norm} of an operator $O\in\B(\H)$ is defined as its largest singular value and is equal to the \emph{operator norm},
\begin{equation}
	\norm{O}_{\infty} = \norm{O} \coloneqq \sup_{|\psi\ket\in\H} \frac{\norm{O|\psi\ket}}{ \norm{|\psi\ket} },
\end{equation}
where $\norm{|\psi\ket}=\sqrt{\bra\psi|\psi\ket}$ denotes the \emph{vector $2$-norm}.
The \emph{$\infty$-norm} is the physically relevant norm for observables.
The \emph{$1$-norm}, of an operator $O\in\B(\H)$ is defined as the sum of its singular values and is equal to the \emph{trace norm},
\begin{equation}
	\norm{O}_1 = \norm{O}_{\operatorname{tr}} \coloneqq \Tr\sqrt{O^\dag O}.
\end{equation}
It is the physically relevant norm for states, i.e., density matrices \cite{Nielsen2000}.
Those operator norms induce corresponding norms for super-operators $T\in \B(\B(\H))$.
The \emph{\InfToInf-norm} is defined as
\begin{equation}
	\norm{T} \coloneqq \norm{T}_{\infty\to\infty} \coloneqq \sup_{O\in\B(\H)} \frac{\norm{T O}_\infty}{\norm{ O }_\infty}
\end{equation}
and the \emph{\OneToOne-norm} is
\begin{equation}
	\norm{T}_{1\to 1} \coloneqq \sup_{O\in\B(\H)} \frac{\norm{T O}_1}{\norm{ O }_1}.
\end{equation}

In order to switch between the Schr\"{o}dinger and the Heisenberg picture, one needs to consider the adjoint $T^\dag$ of a super-operator $T$, defined by
\begin{equation}
	\bra A,TB\ket_\HS=\bra T^\dag A,B\ket_{HS}\quad \forall_{A,B\in \B(\H)},
\end{equation}
where $\bra \cdot,\cdot\ket_\HS$ denotes the Hilbert-Schmidt inner product $\bra A,B\ket_\HS \coloneqq \Tr(A^\dag B)$. 
The \OneToOne-norm is \emph{dual} to the \InfToInf-norm in the sense that
\begin{align}\nonumber
	\norm{T}_{\infty\to\infty}
	&=\sup_{\norm{O}_\infty=1}\norm{TO}_\infty\\\nonumber
	&=\sup_{\norm{O}_\infty=1}\sup_{\norm{X}_1=1}|\bra X,TO\ket_\HS|\\\nonumber
	&=\sup_{\norm{X}_1=1}\sup_{\norm{O}_\infty=1}|\bra T^\dag X,O\ket_\HS|\\\label{eq:normDuality}
	&=\sup_{\norm{X}_1=1}\norm{T^\dag X}_1
	\,\,=\norm{T^\dag}_{1\to 1}.
\end{align}
This allows us to relate the appropriate norm of a propagator $T$ in the Heisenberg picture to the norm of the corresponding propagator $T^\dag$ in the Schr\"{o}dinger picture.
For more on properties of the norms, see, for example, Refs.~\cite{Bhatia1997,Watrous2005-5}.

\section{Properties of the propagators}\label{sec:propagators}
The derivative of a propagator with respect to its second time argument is given by
\begin{equation}\label{eq:propDerivative2}
	\partial_t \tau_V(s,t) = \tau_V(s,t)\L_V(t).
\end{equation}
Using the defining properties $\partial_s \tau_V(s,t) = -\L_V(s) \tau_V(s,t)$ and $\tau_V(t,t)=\Id$, Eq.~\eqref{eq:propDerivative2} follows from the equation
\begin{multline*}
	0 = \partial_t \Id = \partial_t [\tau_V(t,s)\tau_V(s,t)]\\
	 = [\partial_t \tau_V(t,s)]\tau_V(s,t) + \tau_V(t,s)[\partial_t\tau_V(s,t)]
\end{multline*}
after applying $\tau_V(s,t)$ to it.

Let us explain why the propagators are norm-decreasing, i.e., 
\begin{equation}\label{eq:propNormAppx}
	\norm{\tau(s,t)O}\leq \norm{O}\quad \forall\,{\L\in\LL_\Latt,\,s\leq t,\, O\in\B(\H)}.
\end{equation}
The adjoint propagator $\tau^\dag(s,t)$ (see \appx~\ref{sec:norms}), describes the time-evolution in the Schr\"{o}dinger picture, $\dm(t) = \tau^\dag(s,t)\dm(s)$, where $\dm(t)$ denotes the system state at time $t$. 
First of all, we note that $\tau^\dag(s,t)$ is a completely positive, trace-preserving (CPT) map since it can be written as a \emph{product integral} \cite{Dollard1979},
\begin{align*}
\tau^\dag(s,t) = \lim_{\Delta t_j \to 0} \prod_j e^{ 
\L^\dag(t_j) \Delta t_j}.
\end{align*}
Every factor $e^{\L^\dag(t_j) \Delta t_j}$ is an exponential of a constant Liouvillian and is hence CPT. Thus, the finite products are CPT maps and, since the set of CPT maps is closed, also the limit $\tau^\dag(s,t)$ is a CPT map.
Then Eq.~\eqref{eq:propNormAppx} follows from the norm duality $\normS{T}\equiv\normS{T}_{\infty\to\infty}=\normS{T^\dag}_{1\to 1}$ [Eq.~\eqref{eq:normDuality}] and $\normS{T^\dag}_{1\to 1}=1$ for all CPT maps $T^\dag$. The latter has, for example, been shown in Ref.~\cite{Kliesch2011-107}.

\section{Bound on the partial exponential sum}\label{sec:expPartialBound}
In the following, we prove that
\begin{equation}\label{eq:expPartialBound}
	\underbrace{\textstyle\sum_{n=N}^\infty \frac{x^n}{n!} }_{=:f_N(x)}
	\leq
	\underbrace{e^{xe-N}}_{=:g_N(x)} \quad\forall x\geq 0,\,N\in\NN_0.
\end{equation}
Note first that, for $N=0$,
\begin{equation*}
	f_0(x)=e^x\leq e^{xe} =g_0(x)\quad \forall_{x\geq 0}.
\end{equation*}
The functions $f_N$ and $g_N$ obey the differential equations
\begin{equation*}
	\partial_x f_{N+1}(x)=f_N(x),\quad \partial_x g_{N+1}(x)=g_N(x)\quad \forall_{x,N}.
\end{equation*}
For all $N>0$, the initial values $f_N(0)=0$ and $g_N(0)=e^{-N}$ obviously obey
$f_N(0)\leq g_N(0)$ $\forall_{N>0}$.
Consequently, $f_N(x)\leq g_N(x)$ $\forall_{x\geq 0}$ implies $f_{N+1}(x)\leq g_{N+1}(x)$ $\forall_{x\geq 0}$.
This proves Eq.~\eqref{eq:expPartialBound} inductively.

\section{Bound on a sum of exponentials}\label{sec:expBound}
In the following, it is shown that
\begin{equation}\label{eq:expBound}
	\sum_{n=D}^\infty n^\kappa e^{-n} \leq 2e D^\kappa e^{-D}
	\quad\forall \kappa>0,\,D>2\kappa+1\in\NN.
\end{equation}
Due to the definition $\Gamma(a,x) \coloneqq \int_x^\infty \ud t\, t^{a-1} e^{-t}$ of the incomplete Gamma function,
one has
\begin{equation}\label{eq:GammaBoundsSum}
	\sum_{n=D}^\infty n^\kappa e^{-n} \leq \Gamma(\kappa+1,D-1).
\end{equation}
The bound 
\begin{equation*}
	\Gamma(a,x)\leq B x^{a-1}e^{-x}\quad\forall\textstyle a>1,\, B>1,\, x>\frac{B(a-1)}{B-1}
\end{equation*}
of Natalini and Palumbo \cite{Natalini2000-3}, reads for the choice $B=2$
\begin{equation*}
	\Gamma(a,x)\leq 2 x^{a-1}e^{-x}\quad\forall a>1,\, x>2(a-1).
\end{equation*}
Together with Eq.~\eqref{eq:GammaBoundsSum} one obtains
\begin{equation*}
	\sum_{n=D}^\infty n^\kappa e^{-n} \leq 2 (D-1)^\kappa e^{-D+1}
	\quad\forall_{\kappa>0,\,D>2\kappa+1}
\end{equation*}
and hence the acclaimed Eq.~\eqref{eq:expBound}.

\section{Trotter expansion of a propagator}\label{sec:doTrotter}
For two times $q\leq t$, we derive the Trotter error bound 
\begin{multline}\label{eq:doTrotter}
	\normS{\tau_{V}(q,t)
	-\prod_{Z\subset V} \tau_Z(q,t)}_Y\\
	\leq (t-q)^2\Z \vol{V} \normt{\ell}^2 e^{(t-q)\normt{\ell}}
\end{multline}
that is employed in the proof of Theorem~\ref{th:TrotterEvol}.
To this purpose, let us determine an upper bound for the right-hand side of
\begin{equation*}
	\norm{T_{\L+\ell}^{q,t}-T_\L^{q,t}T_\ell^{q,t}}_Y
	\leq \norm{T_{\L+\ell}^{q,t}-T_\L^{q,t}T_\ell^{q,t}},
\end{equation*}
where $T_\K^{q,t}$ denotes the propagator for a Liouvillian $\K(t)\in\LL$, $\L(t)\in\LL$ obeys the preconditions of Theorem~\ref{th:TrotterEvol}, and $\ell(t)\in\LL_Z$ is a local Liouvillian term with support $Z$.
We denote the inverse of a propagator $T_\K^{q,t}$ by $T_\K^{t,q}$.
Using $\partial_q T_\K^{q,t} = -\K(q) T_\K^{q,t}$, $\partial_t T_\K^{q,t} = T_\K^{q,t} \K(t)$, $T_\K^{r,s}T_\K^{s,t}=T_\K^{r,t}$, $T_\K^{t,t}=\Id$, and applying the fundamental theorem of calculus twice, one finds
\begin{align*}
	&T_{\L+\ell}^{q,t} -T_\L^{q,t}T_\ell^{q,t} 
	 = (T_{\L+\ell}^{q,t} T_\ell^{t,q}T_\L^{t,q} -\Id )T_\L^{q,t}T_\ell^{q,t}\\
	&\,\,= \int_q^t\ud s\, \partial_s \big( T_{\L+\ell}^{q,s} T_\ell^{s,q}T_\L^{s,q} \big)
	         T_\L^{q,t}T_\ell^{q,t}\\
	&\,\,= \int_q^t\ud s\, T_{\L+\ell}^{q,s}
	  \big( \L(s) - T_\ell^{s,q} \L(s) T_\ell^{q,s}  \big) T_\ell^{s,q} T_\L^{s,t}T_\ell^{q,t} \\
	&\,\,= \int_q^t\ud s\, \int_q^s\ud r\, T_{\L+\ell}^{q,s} \partial_r 
	  \big(T_\ell^{s,r} \L(s) T_\ell^{r,s} \big) T_\ell^{s,q} T_\L^{s,t}T_\ell^{q,t} \\
	&\,\,= \int_q^t\ud s\, \int_q^s\ud r\, T_{\L+\ell}^{q,s} T_\ell^{s,r} [\ell(r), \L(s)] T_\ell^{r,q} T_\L^{s,t}T_\ell^{q,t}.
\end{align*}
The time arguments occurring in the integrand are ordered according to $q\leq r\leq s\leq t$. The norm of the propagators is $\norm{T_\K^{s,t}}=1$ $\forall_{s\leq t}$. A bound for the norm of the inverse propagators can be obtained from their representations as time-ordered exponentials \cite{Kliesch2011-107,Dollard1979}, yielding $\norm{T_\K^{t,s}}\leq \exp(\int_s^t\ud r\,\norm{\K(t)} )$ $\forall_{s\leq t}$. With those properties, the triangle inequality, and the norm submultiplicativity,
\begin{align*}
	\norm{T_{\L+\ell}^{q,t} -T_\L^{q,t}T_\ell^{q,t}}
	&\leq \int_q^t\ud s\, \int_q^s\ud r\,  \norm{[\ell(r), \L(s)]} e^{(s-q)\normt{\ell}}\\
	&\leq (t-q)^2 \Z  \normt{\ell}^2 e^{(t-q)\normt{\ell}}.
\end{align*}
This bound and the inequality $\normS{T_1T_2-\tilde{T}_1\tilde{T}_2}
\leq \norm{T_1}\normS{T_2-\tilde{T}_2} + \normS{T_1-\tilde{T}_1}\normS{\tilde{T}_2}$ can now be used iteratively, separating one local propagator $T_{\ell_Z}^{q,t}$ after another. As $\L_V$ is a sum of $\vol{V}$ terms $\ell_Z$, Eq.~\eqref{eq:doTrotter} follows.
\vspace*{20em}

\bibliographystyle{prsty} 

\end{document}